\documentclass[runningheads]{llncs}
\usepackage{graphicx}
\usepackage{microtype}
\usepackage[nolist,nohyperlinks]{acronym}
\newacro{SBG}[SBG]{stochastic Bayesian game}
    \newacroindefinite{SBG}{an}{a}
    \newacroplural{SBG}[SBGs]{stochastic Bayesian games}
\newacro{MCTS}[MCTS]{Monte Carlo tree search}
    \newacroindefinite{MCTS}{an}{a}
    \newacroplural{MCTS}[MCTSs]{Monte Carlo tree seraches}
\newacro{UCT}[UCT]{upper confidence tree}
    \newacroindefinite{UCT}{a}{an}
    \newacroplural{UCT}[UCTs]{upper confidence trees}
\newacro{RL}[RL]{reinforcement learning}
    \newacroindefinite{RL}{an}{a}
\newacro{MARL}[MARL]{multi-agent reinforcement learning}
    \newacroindefinite{MARL}{a}{a}
\newacro{NN}[NN]{neural network}
    \newacroindefinite{NN}{an}{a}
    \newacroplural{NN}[NNs]{neural networks}
\newacro{GNN}[GNN]{graph neural network}
    \newacroindefinite{GNN}{a}{a}
    \newacroplural{GNN}[GNNs]{graph neural networks}
\newacro{CNN}[CNN]{convolutional neural network}
    \newacroindefinite{CNN}{a}{a}
    \newacroplural{CNN}[CNNs]{convolutional neural networks}
\newacro{VAE}[VAE]{variational autoencoder}
    \newacroindefinite{VAE}{a}{a}
    \newacroplural{VAE}[VAEs]{variational autoencoders}
 \newacro{LBF}[LBF]{Level-Based Foraging}
    \newacroindefinite{LBF}{a}{a}
 \newacro{AHT}[AHT]{Ad hoc teamwork}
    \newacroindefinite{AHT}{a}{a}
 \newacro{ROS}[ROS]{robot operating system}
    \newacroindefinite{ROS}{a}{a}
    
\usepackage{xcolor}
\usepackage[numbers,sort]{natbib}
\usepackage{amssymb}
\usepackage{comment}
\usepackage{times}

\usepackage{soul}
\usepackage{url}
\usepackage[hidelinks]{hyperref}
\usepackage[utf8]{inputenc}
\usepackage[small]{caption}
\usepackage{graphicx}
\usepackage{amsmath}
\usepackage{booktabs}
\usepackage{comment}
\usepackage{multirow}
\usepackage{lscape}
\usepackage{soul}
\usepackage{enumitem}
\newenvironment{myquote}%
  {\list{}{\leftmargin=0.2in\rightmargin=0.2in}\item[]}%
  {\endlist}
  
\defcitealias{albrechtComparativeEvaluationMAL2012}{Albrecht et~al.}
\defcitealias{liemhetcharatAllocatingTrainingInstances2017}{Liemhetcharat et~al.}
\defcitealias{albrechtGameTheoreticModelBestResponse2013}{Albrecht et~al.}
\defcitealias{chakrabortyCooperatingMarkovianAd2013}{Chakraborty et~al.}
\defcitealias{bowlingCoordinationAdaptationImpromptu2005}{Bowling et~al.}
\newcommand\mycitet[1]{\citetalias{#1}\ \citep{#1}}

\begin{document}
\title{A Survey of Ad Hoc Teamwork Research}
%
%

\author{
Reuth~Mirsky\inst{1,2} \and 
Ignacio~Carlucho\inst{3,}\thanks{Corresponding Author}\and 
Arrasy~Rahman\inst{3}\and
Elliot~Fosong\inst{3} \and
William~Macke\inst{2}\and
Mohan~Sridharan\inst{4} \and 
Peter~Stone\inst{2,5}\and
Stefano~V.~Albrecht\inst{3} 
}

\authorrunning{Mirsky et al.}
%
\institute{
Bar Ilan University, Israel 
\email{mirskyr@cs.biu.ac.il}
\and
The University of Texas at Austin, USA 
\email{\{wmacke, pstone\}@cs.utexas.edu}
\and
The University of Edinburgh, UK \\
\email{\{ignacio.carlucho, arrasy.rahman, e.fosong, s.albrecht\}@ed.ac.uk}
\and
The University of Birmingham, UK 
\email{m.sridharan@bham.ac.uk}
\and
Sony AI, USA
}

\maketitle              
\begin{abstract}
Ad hoc teamwork is the research problem of designing agents that can collaborate with new teammates without prior coordination. This survey makes a two-fold contribution: First, it provides a structured description of the different facets of the ad hoc teamwork problem. Second, it discusses the progress that has been made in the field so far, and identifies the immediate and long-term open problems that need to be addressed in ad hoc teamwork.

\keywords{Ad Hoc Teamwork \and Collaboration Without Prior Coordination
\and Agent Modelling \and Reinforcement Learning \and Zero-Shot Coordination}
\end{abstract}

\section{Introduction}

\ac{AHT} is defined as the problem of developing agents capable of cooperating on the fly with other agents without prior coordination methods, such a shared task and communication protocols or joint training.
Designing an \ac{AHT} agent is a complex problem, but the underlying capabilities are crucial to enabling agents to take on their designated roles in many practical domains. From service robots and care systems to team sports and surveillance, agents need to reason about the best way to collaborate with other agents and people without prior coordination.
 Research in \ac{AHT} has been around for at least 15 years~\citep{rovatsosSocialComplexityReduction2002, bowlingCoordinationAdaptationImpromptu2005}, and it was proposed as a formal challenge by~\citet{stoneAdHocAutonomous2010}:
\begin{myquote}
    ``To create an autonomous agent that is able to efficiently and robustly collaborate with previously unknown teammates on tasks to which they are all individually capable of contributing as team members.''
\end{myquote}
\noindent
Since then, hundreds of papers that include the phrase ``ad hoc teamwork'' have been published (464 according to Google Scholar at the time of writing this paper) and many more address closely related problems under names such as ``zero-shot coordination''
~\citep{bullardExploringZeroshotEmergent2020, huOtherplayZeroshotCoordination2020}. Moreover, much of the work on personalizing agents' interactions with humans can be viewed as instances of \ac{AHT}~\citep{pittir41667}. 

This survey seeks to make a two-fold contribution. First, it defines the \ac{AHT} problem by describing the underlying assumptions (Section~\ref{sec:formulation}), key subtasks (Section~\ref{sec:subtasks}), and the scope of the problem as considered in this paper (Section~\ref{sec:Components}). Second, it surveys the existing work in \ac{AHT} in terms of the solution methods (Section~\ref{sec:Solutions}) and the evaluation domains that have been developed (Section~\ref{Sec:Evaluation}), and discusses the open problems in the field of \ac{AHT} (Section \ref{sec:conclusion}).

\paragraph{Related initiatives.}
Several initiatives over the last decade have contributed to research progress in \ac{AHT}. In particular, between 2014 and 2017, the Multi-Agent Interaction without Prior Coordination (MIPC) workshop series\footnote{\url{https://mipc.inf.ed.ac.uk}} held at AAAI and AAMAS conferences facilitated discussions and presentations in AHT and related topics. The MIPC workshop series was followed by a special journal issue~\citep{albrechtSpecialIssueMultiagent2017} which featured a collection of new research works in \ac{AHT}. Moreover, the RoboCup Drop-in Challenge was introduced to provide a platform to develop and evaluate \ac{AHT} capabilities in the context of soccer-playing robots~\citep{genterThreeYearsRoboCup2017}.
However, to date there is no comprehensive survey on \ac{AHT}. We seek to address this gap in the literature and help foster further research in \ac{AHT}.

\section{Background}
\label{sec:background}

This section provides a basic formulation of the \ac{AHT} problem. It takes the original challenge proposed in~\cite{stoneAdHocAutonomous2010} and describes it in terms of the inputs and outputs, and the underlying assumptions (Section~\ref{sec:formulation}). It then describes the subtasks of the problem based on issues addressed in relevant papers (Section~\ref{sec:subtasks}).

\subsection{Problem Formulation}
\label{sec:formulation}
The \ac{AHT} problem focuses on training an agent to coordinate with an unfamiliar group of teammates without prior coordination. In this work, we refer to the trained agent as the \textbf{learner}. The learner's \textbf{teammates} are assumed to be capable of contributing to the common teamwork task, meaning that they have a set of skills that are useful for the task at hand. Here we describe the inputs, outputs, and the underlying assumptions of this problem.

\paragraph{Input.} The inputs of the \ac{AHT} problem are the teamwork task to be executed, domain knowledge comprising a description of the domain/environment in which the task is to be executed, a (possibly incomplete) list of attributes characterizing each agent (e.g., a set of goals, perception, and action capabilities), a description of the learner's abilities, and a list of teammates. 
The agent attributes' values might differ between each teammate---also see first assumption below---and some teammates might be able to communicate with each other.%

\paragraph{Output.} The output of the problem is the learner, represented by a policy that determines the action this agent should execute in any given state of the domain. Depending on the agent's sensors, actuators, and the available communication channels, this policy can be deterministic or stochastic, static or adaptable, and might include ontic (physical) actions and epistemic (knowledge-producing) actions, which in turn may contain verbal or non-verbal communication.    

\paragraph{Assumptions.} Three key assumptions (i.e., claims or postulates) characterize the \ac{AHT} problem.

\begin{enumerate}
    \item \textbf{No prior coordination.} The learner is expected to cooperate with its teammates when the task begins without any prior opportunities to establish or specify mechanisms for coordination. For example, it is not possible to prespecify the agents' roles or to have a joint training phase for all agents. The learner might know or assume knowledge of a subset of attributes (e.g., current policies, individual goals) of some subset of its teammates. This  knowledge might be acquired from an expert who has had prior interactions with the learner's current teammates, and the assumptions might be the result of generic models or rules based on past interactions in the target domain. The learner's current teammates might or might not be familiar with one another before the current interaction. For example, in drop-in soccer (a spontaneous soccer match where some or all of the team are strangers), a teammate might be perceived to be a good striker because they are fast and the team can work around this assumption even if they have not played with that specific player before. \\
    
    \item \textbf{No control over teammates.} The learner cannot change the properties of the environment, and the teammates' policies and communication protocols; it has to reason and act under the given conditions. We distinguish between \emph{changing the properties} of the environment (e.g. modifying observability level) and \emph{acting in} the environment to change its state (e.g. picking up a box). Similarly, the learner might influence its teammates' actions, but this influence will be in accordance with the pre-defined policy of the teammates. Moreover, teammates' policies may support learning or adaptation, but the learner cannot modify these abilities. Continuing with the soccer example, teammates can learn to work better together with practice, but no teammate can impose their knowledge on the team before the game starts. \\

    \item \textbf{Collaborative.} All agents are assumed to have a common objective, but some teammates might have additional, individual objectives, or even completely different rewards. However, these additional objectives do not conflict with the common task~\citep{groszEvolutionSharedplans1999}. 
    In the drop-in soccer example, different teammates may have incentives in their contract that encourage them to focus on different skills, e.g., goal-scoring rewards for forwards or assist rewards for midfielders. The difference in the individual objectives may result in situations in which an individual agent may seem to be acting contrary to the team reward, but each agent in the team is always acting to achieve the common objective. For example, although passing frequently is considered very important to a team's performance in a soccer game, an individual teammate may choose to dribble forward because of a perceived opportunity to score a goal.

\end{enumerate}

\subsection{Subtasks in Ad Hoc Teamwork}
\label{sec:subtasks}
Based on a survey of the existing literature, we identified four main subtasks that the learner should be able to perform, although much of the existing work only focuses on addressing a  subset of these subtasks.

\paragraph{ST1: Knowledge representation.} 
The learner requires a representation of the domain knowledge. This includes knowledge about the environment (e.g., discrete or continuous, static or dynamic, etc.), its capabilities, and knowledge about potential teammates (e.g., similarity to past teammates, their theory of mind, etc.). These choices influence the solution methods for the other substasks. 
 Most of the attributes characterizing the environment are common to all multi-agent problems. They can be presented in the classical PEAS system~\citep{russellArtificialIntelligenceModern2021} and are not unique to \ac{AHT}, so we do not elaborate on these here.

\paragraph{ST2: Modeling teammates.} The learner can leverage information about its teammates to improve its decision making. Thus, a key subtask for the learner is to model the information pertaining to teammates' behavior  (e.g., classifying teammates by type in order to adapt to different teammates).

\paragraph{ST3: Action selection.}
The third subtask is
the design of mechanisms used by the learner to select actions once it has an estimate of its teammates' behavior (observed or based on models of teammates). Example methods for this subtask include planning methods and expert policies that are learned or based on expert knowledge.

\paragraph{ST4: Adapting to changes.} 
During interaction, the learner might receive new information about its teammates, the environment, or task objectives. Based on this information, the learner needs to adapt its behavior to improve coordination. This adaptation also includes merging the models provided by teammates.

\section{Boundaries of Ad Hoc Teamwork}
\label{sec:Components}

Here we further define the scope of the \ac{AHT} problem by describing factors that can be considered within the basic problem formulation presented above, and by discussing related research problems.

\subsection{Variations of the Ad Hoc Teamwork Problem}
\label{sec:binary}
We first describe additional factors that define the scope of \ac{AHT} and influence the subtasks described earlier.

\paragraph{Partial observability.} 
Under conditions of full observability, each agent is aware of the state of the environment, including the location of other agents. Partial observability implies a higher level of complexity in knowledge representation as it introduces uncertainty in certain parts of the domain state. Changing the observability level will affect ST1 and thus the other subtasks described above.

\paragraph{Open environment.} 
Closed environments assume a fixed number of teammates~\citep{rahmanOpenAdHoc2021}. Relaxing this assumption increases the problem complexity, as the learner will also have to adapt to the changing number of teammates in the environment; this will primarily affect ST2 and ST4. 

\paragraph{Communication.}  
Since the exploration of how communication can be leveraged to improve team performance is an important area of research in \ac{AHT}, we make a distinction based on whether there is any communication channel between agents. When communication exists, it is sometimes presented as predetermined and known protocols, such as the hints allowed in the game of Hanabi~\citep{bardHanabiChallengeNew2020}, which affects ST1. If these protocols are unknown in the beginning of the interaction and need to be learned during the task execution, it has an effect on ST3 and ST4.

\paragraph{Adaptive teammates.}
We make a distinction between work where the teammates learn alongside the learner, or use policies that stay fixed throughout the learning phase of the learner. Unlike multi-agent reinforcement learning (see Section \ref{sec:related}), which supports joint training for all agents in the team, \ac{AHT} does not assume that the deployed teammates are the same as those the learner might have trained with.  
Rather, adaptive teammates learn by reacting to the learner's policy using methods that are not known to the learner, thus affecting ST3 and ST4. An example of such a setup is flocking, where the teammates have a fixed policy, but their actions are directly influenced by the learner \citep{genterAddingInfluencingAgents2016}.

\paragraph{Mixed objectives.}
While teammates are assumed to be collaborative, they can have mixed objectives. Two types of scenarios arise depending on the objectives of the learner and its teammates. In the first, the learner and the teammates have a perfectly aligned objective (e.g., the reward functions of all agents are identical). In the second, while all team members have a common goal, each agent might also hold individual goals as long as these are not purely adversarial to the shared one. This factor extends the original formulation in~\citep{stoneAdHocAutonomous2010}, is related to the third assumption in Section~\ref{sec:formulation}, and will primarily affect ST2 and ST3.

\subsection{Related Problems}
\label{sec:related}
In this section, we highlight the main differences between \ac{AHT} and other related research problems.

\paragraph{Multi-agent reinforcement learning (MARL).} It refers to the use of reinforcement learning methods for jointly training multiple agents to maximize their respective cumulative rewards while working with each other~\citep{busoniuComprehensiveSurveyMultiagent2008,devlin_kudenko_2016,Papoudakis2019DealingWN}.
\ac{AHT}, on the other hand, assumes control over a single agent (the learner) while teammates can have their own learning mechanisms, e.g., a robot interacting with different human. Prior work has shown that the good team performance of MARL methods often comes at the expense of poor performance when interacting with previously unseen teammates~\citep{vezhnevetsOPtionsREsponsesGrounding2020, rahmanOpenAdHoc2021, huOtherplayZeroshotCoordination2020}. MARL methods are thus not particularly well-suited to \ac{AHT}. 

\paragraph{Ad hoc teaming.}
The objective is to learn coercive measures that may allow self-interested agents with different skills and preferences to collaborate and solve a task. For example, existing work has trained a manager to assign subtasks to agents based on their skills while also incentivizing agents to complete their tasks~\citep{shuRLMindawareMultiagent2018}. In contrast, the learner in \ac{AHT} might incentivize its teammates to act in a certain way, but cannot dictate the teammates' behavior due to the lack of prior coordination.

\paragraph{Agent modelling.}
These methods infer attributes of teammates' behavior such as beliefs, goals, and actions~\citep{albrechtAutonomousAgentsModelling2018}. Since inferring teammates' behavior is important for decision making in \ac{AHT} (e.g., ST3 in Section~\ref{sec:subtasks}), agent modeling methods are  
useful for \ac{AHT}. However, they can be used for a broader class of problems and are not limited to (or necessarily indicative of) \ac{AHT}.

\paragraph{Human-agent interaction.}
The task of creating agents that interact with previously unseen agents has also been explored in the human-agent/robot interaction community. In human-agent interaction, agents have to achieve their goals in the presence of human decision makers. As in \ac{AHT}, it is often impossible to jointly train humans and agents to coordinate their behavior; agents must instead find a way to coordinate with previously unseen humans,  
e.g., by using implicit communication or acting in a legible manner~\citep{breazeal2005effects, dragan2013legibility}.

 \paragraph{Zero-shot coordination (ZSC).} A special case of \ac{AHT} where teammates' behavior are assumed to arise from a reward function that always provides identical rewards for every agent is known as ZSC~\citep{ lupuTrajectoryDiversityZeroshot2021,huOffbeliefLearning2021,bullardExploringZeroshotEmergent2020,bullardQuasiequivalenceDiscoveryZeroshot2021}. After training different populations of agents under the same fully cooperative setup, a ZSC agent is evaluated by measuring its performance when cooperating with agents from a different population. 
 While ZSC introduced techniques relevant for \ac{AHT}, there are \ac{AHT} problems where the controlled agent must interact with teammates whose reward functions are different from its own.

\section{Solution Approaches}
\label{sec:Solutions}
As stated earlier, while existing methods for \ac{AHT} often provide a functioning learner, each method's key contribution can often be mapped to one or more of the four subtasks in Section~\ref{sec:subtasks}. Here we elaborate on common solution methods for each subtask and refer to representative literature.

\subsection{Knowledge Representation}
The representation of domain knowledge strongly influences the solution approach used in the other subtasks. This information can be acquired from human experts (or expert knowledge), prior knowledge of past teammates, or using self-play.
To support adaptation based on limited information, it is common to equip agents with preconceptions of the likely behaviors or intentions of previously unseen teammates. These preconceptions are based on prior experience with the task; this can be the agent's own experience or that of a human familiar with the task. \textit{Agent modeling} techniques can be used to represent the teammates~\citep{albrechtAutonomousAgentsModelling2018}.
  
\paragraph{Type-based methods.}
The use of type-based methods is common in the \ac{AHT} literature. These methods represent prior experience with agents (in the target domain) by a set of hypothesized \emph{types}, where each type models an action selection policy. It is assumed that new teammates encountered by the learner have behaviors specified by one of these types. %
  
  A range of type representations have been explored. Early work explored a nested representation of agents' beliefs, where agents perform Bayesian updates to maintain beliefs over physical states of the environment and over models of other agents~\citep{gmytrasiewiczFrameworkSequentialPlanning2005}. It was also common to use hand-coded programs to represent types~\citep{barrettEmpiricalEvaluationAd2011,albrechtGameTheoreticModelBestResponse2013}. For approaches that employ a learned type set, learned decision trees were a common representation~\citep{barrettMakingFriendsFly2017}. More recently, latent type methods have been used which learn a neural network-based encoder to map observations of teammates to an embedding of the agent's type~\citep{rabinowitzMachineTheoryMind2018,xieLearningLatentRepresentations2020,rahmanOpenAdHoc2021,zintgrafDeepInteractiveBayesian2021}.

  There are three main approaches to specifying a hypothesized type space: (1) specification by a human expert; (2) learning from data; and (3) using \ac{RL} methods and access to the environment or an environment model.   %
  \cite{barrettMakingFriendsFly2017} collect diverse behaviors by drawing their types from the output of an assignment presented to a large number of student. %
  Many methods attempt to generate diverse behaviors in a population trained via \ac{RL}, requiring only access to the target task. 
  They do so using methods such as genetic algorithms~\citep{albrecht2015empirical,albrechtEHBAUsingAction2015,canaanGeneratingAdaptingDiverse2020}, regularisation techniques~\citep{lupuTrajectoryDiversityZeroshot2021}, and reward-shaping techniques~\citep{leiboScalableEvaluationMultiagent2021}.

\paragraph{Experience replay.}
Rather than encoding experience in explicit behavioural models, experience replay methods store transition data in a buffer.  Transitions observed during an interaction are compared against the stored transitions to identify the current teammate~\citep{chenAATEAMAchievingAd2020}.

\paragraph{Task recognition.}
In methods based on task recognition, prior experience or information provided by an expert is encoded as a library of tasks referred to as \emph{plays}, \emph{macro actions}, or \emph{options}~\citep{suttonMDPsSemiMDPsFramework1999}.
Tasks then encode prior experience as applicability conditions, termination conditions, and high-level specifications of a sequence of low-level actions~\citep{wangTooManyCooks2021}.

\subsection{Identifying Current Teammates}
Once a representation is set, estimating the behavior of current teammates allows the learner to determine a suitable behavior.

  \paragraph{Type inference.}
  Methods that represent teammates using types infer beliefs over the hypothesized type space using a history of interactions of the learner with each teammate up to the current timestep. The dominant approach is to use a Bayesian belief update~\citep{albrechtBeliefTruthHypothesised2016,barrettMakingFriendsFly2017}. 
  In such methods, prior beliefs about the teammates' types are updated using the history of interactions and a likelihood of the types based on the history. It is also common to assume uniform priors across types and type parameters~\citep{albrecht2015empirical}. 
  \paragraph{Experience recognition.}\label{sec:experience-recognition}
  Rather than inferring types, some approaches attempt to measure the similarity of the current observations to that from earlier experience in a more direct manner.
  PLASTIC-Policy \citep{barrettMakingFriendsFly2017} compares the most recently observed state transition  to previously stored data. 
 For each team they find the stored transition with the closest state to the current state, and consider the next state observed in that historical transition. They then measure the distance between that state and the observed next state, and use this to compute the likelihood of the team.
  AATEAM \citep{chenAATEAMAchievingAd2020} takes a more sophisticated approach which uses prior experience buffers to train one attention-based neural network per type, to identify agents from a trajectory rather than a single transition.
  
 \paragraph{Task recognition.}
  For methods which represent prior knowledge as tasks, the learner attempts to infer the current task being carried out by the teammate under consideration. \citet{wangTooManyCooks2021} achieved this by assuming that  the teammate was attempting to complete hypothesized tasks and computing the extent to which the teammate's observed behavior is sub-optimal for that task.
  \cite{meloAdHocTeamwork2016} consider a setting in which agents both identify the current task and identify the teammate's strategy, with the teammate's behavior subject to a bounded rationality assumption.
 
\subsection{Action Selection}
Given current knowledge about task and teammates, agents must decide which action to take to maximize team return.

  \paragraph{Planning.}
  Many \ac{AHT} approaches use planning methods to select actions.
  Some, such as \citet{bowlingCoordinationAdaptationImpromptu2005} and \citet{ravulaAdHocTeamwork2019}, use bespoke planning methods suited to the specific task, and chosen by a human expert.
  Many approaches use the more general \ac{MCTS} planning procedure
  \citep{
  wuOnlinePlanningAd2011,
  barrettCommunicatingUnknownTeammates2014,
  alfordActiveBehaviorRecognition2015,
  sarrattTuningBeliefRevision2015,
  albrechtReasoningHypotheticalAgent2017, 
  malikEfficientGeneralizedBellman2018,
  yourdshahiLargeScaleAdhoc2018,
  eckScalableDecisiontheoreticPlanning2020}.
  The \ac{UCT} algorithm \citep{UCTAlgorithm} for \ac{MCTS} is often used due to its ability to perform well when the branching factor is large, as is the case when multiple agents are present.
  These \ac{MCTS}-based methods require that types are represented by explicit behavioral models to sample teammate actions during rollouts.
  
  \paragraph{Expert policy methods.}
  Selecting actions by choosing a policy from a set of expert policies, and then acting according to the chosen policy.
  There are many ways in which these expert policies can be obtained prior to the ad hoc interaction: they can be provided by an expert, 
  learned offline, using experience data \citep{chenAATEAMAchievingAd2020, santosAdHocTeamwork2021},
  or by online \ac{RL} training given the task \citep{albrechtEHBAUsingAction2015}. %
  One of the advantages of expert policy methods over type-based planning methods is that they can handle large or continuous state and action spaces, where \ac{MCTS} approaches may struggle \citep{barrettMakingFriendsFly2017}.
  However, type-based planning methods are more appropriate when the ad hoc team is likely to have a previously unseen composition, as type-based methods can reason at the level of the types of individual agents. Also, creating expert policies may be impossible when a large variation of situations are encountered.
  The E-HBA method attempts to achieve the advantages of both type-based reasoning methods and expert policy methods by combining the two \citep{albrechtEHBAUsingAction2015}.
  The GPL method \citep{rahmanOpenAdHoc2021}, suitable in open \ac{AHT} problems, 
   uses an action-selection mechanism based on E-HBA .

  \paragraph{Leading.}
  Some works explicitly consider adaptive teammates, where a learner's choice of action affects its teammates' behaviors. Works such as \citet{agmonModelingUncertaintyLeading2014}
  assume teammates employ a known best response strategy, and that the goal is to lead these teammates to a specific joint coordination strategy.
  These approaches were addressed in simple games using dynamic programming.
  \Citet{xieLearningLatentRepresentations2020}
  consider cases where the learner does not know the teammate's current behavior, nor how this behavior changes across interactions. Thus, deep learning is used to learn an embedding of the teammate's strategy, and model the teammate's behavioral dynamics and teammates' adaptation process.

  \paragraph{Metalearning.}
  Metalearning approaches use action selection policies which are trained to facilitate the entire \ac{AHT} process. The MeLIBA approach \citep{zintgrafDeepInteractiveBayesian2021} trains the policy to carry out interactive Bayesian \ac{RL}, 
  intentionally taking actions which seek to reveal information about the teammate's type.
  The action selection policies of metalearning approaches is typically conditioned on the learner's prediction of the teammate's type.
  In this sense, such methods can be compared to expert policy methods.
 
\subsection{Adapting to Current Teammates}
 
  During interaction, the learner receives new information, which can be used to adapt its behavior. 
  
  \paragraph{Belief revision.}
   Most methods employ belief revision protocols to maintain their belief about the identity of other agents across time.
   For type-based methods, it is typical to assume each teammate's type does not change over time, and that a good representation of the teammate exists in the hypothesized type space \citep{albrechtBeliefTruthHypothesised2016}.
   However, if it is assumed that teammates' types change over time, the learner must also adapt.
   The ConvCPD method \citep{ravulaAdHocTeamwork2019} considers settings in which the type space is known, but agents can switch types. For these settings, they employ a \ac{CNN}-based changepoint detection approach, which uses image-like representations of type likelihoods across time to detect changes. %
   An alternative approach is to modify the Bayesian belief revision process to allow beliefs to decay towards the priors over time. This approach is useful when a teammate changes to a type which the learner has assigned low (or zero) probability to. In this case, the learner might struggle (or be unable) to quickly update its belief to reflect the new true teammate \citep{ santosAdHocTeamwork2021}. Sum-based posterior definitions were also proposed to deal with changing types \citep{albrechtBeliefTruthHypothesised2016}. 
  
 \paragraph{Hypothesis space revision.}
  Approaches exist for adapting to agents whose behavior may not be adequately represented in the hypothesized space.
  TwoStageTransfer is a transfer learning method employed by PLASTIC-Model \citep{barrettMakingFriendsFly2017} which uses observations of new teammates and prior  models to finetune a model for the new teammate.

  \paragraph{Metalearning.}
  During the metalearning process, the action selection policy learns its own adaptation procedures, avoiding the need to specify particular adaptation schemes \citep{xieLearningLatentRepresentations2020, zintgrafDeepInteractiveBayesian2021}.

  \paragraph{Zero-Shot coordination techniques.}
  The ZSC problem does not allow the learner any behavioral adaptation during ad hoc interactions.
  For this reason, the focus of these methods is on training agents which robustly coordinate with other agents trained using the same algorithm.
  One approach is to avoid strategies which are not invariant under symmetries within the underlying tasks \citep{huOtherplayZeroshotCoordination2020,huOffbeliefLearning2021}.
  Another approach is based on the hypothesis that there are few strategies which perform well with a diverse set of teammates, so ad hoc agents independently trained against diverse teammates (and themselves) are likely arrive at similar pre-coordinated policies \citep{lupuTrajectoryDiversityZeroshot2021}. 

 \paragraph{Communication.} The learner can quickly adapt to changes is by communicating with its teammates. This communication can either be a query \citep{mirskyPennyYourThoughts2020,mackeExpectedValueCommunication2021}, transfer knowledge or preferences \citep{meadImpromptuTeamsHeterogeneous2007, barrettCommunicatingUnknownTeammates2014}, or providing an advice \citep{shvo2020active, canaanGeneratingAdaptingDiverse2020}.

\section{Evaluation Domains}
 \label{Sec:Evaluation}

Many different approaches have been used for evaluating \ac{AHT} methods. In this section, we categorize them using the identified variations from Subsection~\ref{sec:binary}. Some domains might fit more than one category, but we place them according to the first ad hoc teamwork paper they appeared in. In Table~\ref{tab:domains}, we summarize each of the domains and associated papers.

  \paragraph{No variations.}
  Some evaluation domains do not have any of the variations outlined in Section~\ref{sec:binary}. Among these \ac{AHT} domains, some of the simplest are matrix games~\citep{albrechtEHBAUsingAction2015,meloAdHocTeamwork2016}. These games consist of a payoff matrix for two agents who independently choose actions and then receive a payoff based on the actions each agent chose. The game is then repeated with the goal to maximize long term return over repeated trials.
  Another common domain is predator prey \citep{barrettEmpiricalEvaluationAd2011,ravulaAdHocTeamwork2019, papoudakisLocalInformationAgent2021}. This domain consists of several agents (the predators) attempting to surround and capture other agents (the prey). The predator prey domain requires both recognising a teammate's goal (namely which prey they are pursuing), and also collaborating with other agents to surround the prey.
  In level-based foraging \citep{albrechtGameTheoreticModelBestResponse2013}, the goal of the agent team is to collect food items which are spatially distributed in a grid world. Agents and items have different skill levels which represent different capabilities in agents, requiring that agents decide when and with whom to collaborate in order to collect the items.
  
   \paragraph{Open environments.} 
  There are several instances of open domains presented in \ac{AHT}. First, open variations of the domains mentioned above exist in \citet{rahmanOpenAdHoc2021}. 
  Another open \ac{AHT} domain is wildfires, where agents entering and leaving the environment need to work together to contain the spread of wildfires \citep{chandrasekaranIndividualPlanningOpen2016}.
  Finally, ad hoc flocking and swarming domains enable agents to enter and leave the environment freely~\citep{genterAddingInfluencingAgents2016}.
  
    
   \paragraph{Partial or noisy observability.}
  Partially observable variants of the domains with no extensions exist in \citet{ribeiroAssistingUnknownTeammates2022}.
  One domain that has been prevalent in \ac{AHT} literature is robot soccer. Drop-in soccer where a group of players need to form a team without playing with each other is common among humans in real life, so it has been a frequented challenge by AI as well \citep{barrettMakingFriendsFly2017, genterThreeYearsRoboCup2017}. The problem typically consists of substituting one member of a team with a learner. The performance is then measured on how robust the learner's performance is regardless of which team it is placed in.
  This domain presents an additional challenge, as each agent can only observe its local environment. 
  Another partially observable domains are military simulation, which simulate various combat and search tasks using unmmaned autonomous vehicles 
  \citep{alfordActiveBehaviorRecognition2015}, and the collaborative card game Hanabi \citep{bardHanabiChallengeNew2020}. Similar to the RoboCup domain, these domains also present the challenge that agents only have access to their local observations.

 \begin{table}[]
 \small
     \centering
      \caption{Different environments used for evaluating ad hoc teamwork. } 
     \label{tab:domains}
     \begin{tabular}{|p{1.4cm}|l|p{6.9cm}|}
     \hline
     \textbf{Domain}&\textbf{Paper}&\textbf{Method Description} \\
     \hline
          {\multirow{10}{=}{Matrix Games}}
          &\mycitet{albrechtComparativeEvaluationMAL2012}&Empirically evaluates various multi-agent learning algorithms in ad hoc mixed teams. \\ \cline{2-3} 
          &\mycitet{chakrabortyCooperatingMarkovianAd2013}
          &Introduces an optimal algorithm to cooperate with a Markovian teammate. \\ \cline{2-3} 
          &\citet{albrechtEHBAUsingAction2015}& Combines type-based reasoning for prediction with expert algorithms for decision making. \\  \cline{2-3} 
            &\citet{albrechtBeliefTruthHypothesised2016,albrecht2015empirical} & Evaluates impact of prior beliefs in type-based reasoning in a range of matrix games. \\  \cline{2-3} 
          &\citet{meloAdHocTeamwork2016}&Extends ad hoc teamwork to scenarios where the current task is unknown in addition to the teammates. \\  \cline{2-3} 
          \hline
          {\multirow{6}{=}{Predator Prey}}
          &\citet{barrettEmpiricalEvaluationAd2011}& MCTS (UCT) with type-based reasoning using hand-crafted types in the predator prey domain. \\  \cline{2-3} 
          &\citet{ravulaAdHocTeamwork2019}&Extends ad hoc teamwork methods  to work with teammates which can switch behaviors. \\  \cline{2-3} 
          &\citet{papoudakisLocalInformationAgent2021}&Assumes only local observations of ad hoc teamwork agent are available to model other agents. \\  \cline{2-3} 
          \hline
          {\multirow{6}{=}{LBF}} 
          &\mycitet{albrechtGameTheoreticModelBestResponse2013}
          & Develops type-based reasoning based on game theory model to solve ad hoc teamwork problems. \\  \cline{2-3} 
          &\mycitet{liemhetcharatAllocatingTrainingInstances2017} \quad
          &Defines the problem of ad hoc team assignment. \\  \cline{2-3} 
          &\citet{yourdshahiLargeScaleAdhoc2018}&Introduces new history-based MCTS. \\  \cline{2-3} 
          &\citet{rahmanOpenAdHoc2021}&Uses graph-based learning to handle   a dynamic number of agents in the environment. \\  
          \hline
           {\multirow{2}{=}{Wildfires}} &\citet{eckScalableDecisiontheoreticPlanning2020}&Introduces ad hoc teamwork in open environments with large numbers of agents. \\
          \hline
          {\multirow{4}{=}{Flocking  Swarming}} &\citet{genterInfluencingFlockAd2014}&Introduces AHT approaches for influencing a flock's behavior. \\  \cline{2-3} 
          &\citet{genterDeterminingPlacementsInfluencing2015}&Determines where to place agents in a flock. \\  \cline{2-3} 
          &\citet{genterAddingInfluencingAgents2016}&Solves how to force agents to join flock in motion. \\
          \hline
          {\multirow{6}{=}{Robot Soccer}} 
          &\mycitet{bowlingCoordinationAdaptationImpromptu2005}&Introduces two new approaches for working with ad hoc teams in robot soccer.  \\  \cline{2-3} 
          &\citet{barrettCooperatingUnknownTeammates2014}&Introduces new method for reusing policies learned from previous teammates to accomplish AHT. \\  \cline{2-3} 
          &\citet{barrettMakingFriendsFly2017}&Introduces algorithms for AHT based on previously met teammates, using either policies or models. \\  
          \hline
          {\multirow{2}{=}{Military Simulation}} & \citet{alfordActiveBehaviorRecognition2015}&Introduces an algorithm for classifying agent behaviors in air combat simulator. \\ 
          \hline
          {\multirow{10}{=}{Hanabi}}&\citet{bardHanabiChallengeNew2020}&Proposes the Hanabi game as a new challenge for AI research, including ad hoc teamwork. \\  \cline{2-3} 
          &\citet{canaanGeneratingAdaptingDiverse2020}&Creates a meta-strategy for solving ad hoc teamwork in Hanabi using a diverse set of possible teammates. \\  \cline{2-3} 
          &\citet{huOtherplayZeroshotCoordination2020}&An effective algorithm for learning from self-play by attempting to seek out new behaviors. \\  \cline{2-3} 
          &\citet{huOffbeliefLearning2021}&Introduces improved method off-belief learning for learning from self-play in DecPOMDPs. \\  \cline{2-3} 
          &\citet{lupuTrajectoryDiversityZeroshot2021}&Creates a new optimisable metric for determining policy diversity in Hanabi self-play. \\  
          \hline
          {\multirow{5}{=}{Tool Fetching Domain}}&\citet{mirskyPennyYourThoughts2020}&Introduces SOMALI CAT problem and proposes solution for determining when queries might be useful. \\  \cline{2-3} 
          &\citet{mackeExpectedValueCommunication2021}&Proposes a solution for what to query when multiple possible queries are available. \\  \cline{2-3} 
          &\citet{suriadinataReasoningHumanBehavior2021}&Investigates human behavior in the Tool Fetch Domain. \\
          \hline
     \end{tabular}
 \end{table}

   \paragraph{Communication.}
  Multiple domains allow communication in some form. The RoboCup domain mentioned above allows limited communication between agents using wireless connections. Others use communication as a more critical part of the domain. 
  The tool fetching domain provides an \ac{AHT} domain that allows one agent to query another about its goals \citep{mackeExpectedValueCommunication2021}. Unlike other domains mentioned so far, the tool fetching domain is specifically focused on evaluating an agent's ability to communicate effectively.
    The Hanabi domain also presents a structured communication channel. While in the tool fetching domain the learner can query its teammates, in Hanabi the communication channel allows the learner to provide its teammates with information unknown to them \citep{bardHanabiChallengeNew2020,canaanGeneratingAdaptingDiverse2020}.
  Another domain that focuses on communication is the cops and robbers domain \citep{sarrattTuningBeliefRevision2015}. In this domain, teammates (cops) must work together to capture another, adversarial agent (the robber). Each agent can query the other to gain information about their current plans \citep{sarrattTuningBeliefRevision2015}.
  %
  
  \paragraph{Adaptive teammates.}
 So far all domains mentioned are focused on evaluating whether a learner can successfully adapt their behavior to collaborate with diverse teammates. Some domains, however, instead try to evaluate how well learner(s) can influence other agents to achieve better performance. While the above domains can be adapted to have learning teammates, several domains exist with this explicit purpose in mind.
 Some examples of these are domains focused on incentivising the teammate to take a specific course of action \citep{wangTooManyCooks2021}, or on swarming \citep{genterInfluencingFlockAd2014}, where the learner attempts to move in such a way as to influence the overall behavior of the agents around it. 
 
   \paragraph{Mixed objectives.}
  Works that make the assumption of coupled objectives, such as ZSC~\citep{huOtherplayZeroshotCoordination2020}, utilize an environment in which the reward received by all agents is the same. Such environments include the lever environment \citep{huOtherplayZeroshotCoordination2020} and Hanabi~ \citep{bardHanabiChallengeNew2020}. 
  Works which do not assume coupled objectives utilize general-sum domains such as level-based foraging \citep{albrechtGameTheoreticModelBestResponse2013}, in which the reward changes depending of the contribution of the agent; or the tool fetching domain where each agent has a distinct role in the team \citep{mirskyPennyYourThoughts2020}.

\section{Conclusion and Open Problems}
\label{sec:conclusion}
In this survey, we presented a review of the \ac{AHT} literature that has been published over the past decade. This long period of time, along with the abundance of published work, enabled us to draw a big picture view of this topic: setting the boundary on what is, and what is not, \ac{AHT}; identifying the subtasks that an agent needs to tackle as part of an \ac{AHT} task; and the various levels of complexity in \ac{AHT}. 
Many open problems still need to be addressed to achieve a robust agent that is able to interact with teammates without prior coordination and solve real-world problems. Furthermore, AHT research is currently suffering from a lack of standardised comparison between existing AHT approaches, which increases the difficulty of identifying state-of-the-art methods for solving a certain AHT problem.

Future work could address further extensions of the variations of the ad hoc teamwork problem discussed in Section \ref{sec:binary}, or combinations of these variations. For example, considering the presence of teammates with complex adaptive processes, such as teammates which learn via RL while interacting with the learner; or teammates which themselves apply AHT techniques. Current approaches to AHT are not designed to work with adaptive teammates (one notable exception being HBA \citep{albrechtBeliefTruthHypothesised2016}), whose presence would mean that the learner needs not only to adapt to teammates' behaviors, but also consider how the teammates adapt to its own behavior.
Another extension is the combination of partial observability and open teams, which provides a difficult challenge for the learner, due to this complex dual uncertainty.

In terms of potential solution methods, one of the crucial open problems is improving the generalization to new teammates that have not yet been seen during training. 
Recent continual learning \citep{PrecupCRL2020} advances showed that training on diverse tasks can result in agents with robust performance in previously unseen tasks \citep{OpenEndedLearning}. 
In the same way, training with a diverse set of teammates can improve the learner's ability to collaborate with new teammates. 
\citet{lupuTrajectoryDiversityZeroshot2021} proposed a method to generate diverse teammates for ZSC, but it was not evaluated with collaborative teammates with objectives that might not be fully aligned with the learner's. Recently, \citet{Rahman2022TG} proposed a method for generating a diverse set teammates specifically for ad hoc teamwork applications. However, results were only obtained in a $5x5$ grid world environment, more work is needed to evaluate how this method performs in more complex environments. These works are a good starting point when designing learners that are robust to different teams, however, they do not specifically address the collaborative aspect of AHT. Additional work is required to properly define the scope of the diverse set of agents a learner should be able to work with.
And while generating teammates that display different behaviours and skill levels can improve generalisation during execution time, this is not an easy task, especially in more complex domains. 

\ac{AHT} research could also benefit from the use of more complex or realistic
domains in evaluation. Previous works tended to use simple domains (Section~\ref{Sec:Evaluation}), but these solutions might not perform well in realistic domains.
We suggest that future AHT research should consider more realistic testbeds, which can rely on robotics simulators extended to handle multi-agent scenarios~\citep{collins2021review},
or on existing scenarios such as the DARPA ``Spectrum Collaboration Challenge''\footnote{\href{https://www.darpa.mil/program/spectrum-collaboration-challenge}{www.darpa.mil/program/spectrum-collaboration-challenge}}, 
which will allow for the evaluation of more complex tasks and algorithms. 
Social navigation, the problem of a robot navigating through a crowd of people and robots, is another relevant robotics challenge \citep{mirsky2021prevention}. 
In this problem, the learner needs to coordinate with previously unmet passerby humans and robots in order to avoid collisions, while allowing each other to get to their destinations. Thus, this challenge poses a series of challenging \ac{AHT} problems where the learner need to adapt to new incoming teammates based on a highly limited amount of interaction experience.

Another important issue that can be addressed by future work is benchmarking current AHT approaches by providing systematic comparison between them. Existing works in AHT often forgo comparison against other approaches designed to solve the same variation of AHT problems, which makes it hard to identify state-of-the-art approaches in the field. A systematic benchmark between AHT approaches across different environments could therefore be a crucial stepping stone towards further identifying the strengths and weaknesses of different AHT methods.

To conclude, the \ac{AHT} problem comprises a unique mixture of subtasks that the learner is required to perform, which requires solutions ranging from different fields.
In this survey, we identified the existing and open problems in \ac{AHT} which we hope will contribute to the development of the field, and in turn will advance the multi-agent research community as a whole.


\bibliographystyle{plainnat}
\bibliography{ijcai22_too_much} 

\end{document}